# THz and infrared studies of multiferroic hexagonal $Y_{1-x}Eu_xMnO_3$ (x=0 - 0.2) ceramics

V. Goian, S. Kamba, C. Kadlec, D. Nuzhnyy, P. Kužel,

Institute of Physics, ASCR, Na Slovance 2, 18221 Prague 8, Czech Republic

J. Agostinho Moreira, A. Almeida

Departamento de Física da Faculdade de Ciências, IFIMUP, Universidade do Porto, Rua do Campo Alegre, 687, 4169-007, Porto, Portugal

P.B. Tavares

Centro de Química de Vila Real, Universidade de Trás-os-Montes e Alto Douro. Apartado 1013, 5001-801, Vila Real, Portugal

**Abstract**

We report an investigation of hexagonal $Y_{1-x}Eu_xMnO_3$ ceramics with x=0, 0.1 and 0.2 using infrared and THz spectroscopies in the temperature range between 5 and 900 K. The temperature dependence of the THz permittivity reveals a kink near the antiferromagnetic phase transition temperature $T_N \approx$ 70 K giving evidence of a strong spin-lattice coupling. Below $T_N$ two absorption peaks were revealed in the THz spectra close to 43 and 73 cm$^{-1}$. While the first peak corresponds to a sharp antiferromagnetic resonance exhibiting softening on heating towards $T_N$, the second one may be attributed to an impurity mode or a multiphonon absorption peak. High-temperature THz spectra measured up to 900 K reveal only small gradual increase of the permittivity in agreement with a weak phonon softening observed in the infrared reflectance spectra upon heating. This corresponds to an improper ferroelectric character of the phase transition proposed from first principle calculations by Fennie and Rabe [Phys. Rev. B 72 (2005), pp. 100103(R)].

**Introduction**

Multiferroic rare-earth manganites $RMnO_3$ (R=rare earth) are intensively studied materials for their pronounced coupling between magnetization and ferroelectric polarization (magnetoelectric effect) [1] which can be used in technological applications like non-volatile memories [2], magnetic sensors or ferroelectric-gate field-effect transistors [3]. The rare-earth manganites exist in two crystallographic forms: manganites with a relatively small rare-earth ionic radius crystallize in a hexagonal structure, while manganites with a larger ionic radius of the rare earth elements are orthorhombic with $MnO_6$ octahedra forming the perovskite structure. The hexagonal manganites are characterized by $MnO_5$ bipyramids which form layeres in the *a-b* plane. Along the *c* direction the layers of $MnO_5$ are well separated by rare earth elements [4].

One of the most intensively studied manganites is hexagonal $YMnO_3$, although yttrium does not belong to rare earth ions. $YMnO_3$ is a high-temperature ferroelectric ($T_C$ near 940 K) [5] and it becomes antiferromagnetic (AFM) below 70 K [6,7]. It crystallizes in the space group $P6_3/mmc$ ($D_{6h}^4$, Z=2) in the high temperature paraelectric phase [8] and it transforms to the polar space group $P6_3cm$ ($C_{6v}^3$, Z=6) in the ferroelectric phase below ~ 940 K [9]. The absence of large Born effective charges and the polarization obtained by a buckling of layered $MnO_5$ bipyramids accompanied by off-centered displacements of Y and in-plane O ions has been referred to a geometrical ferroelectricity [4]. The polarization was found to be rather high, around 5 $\mu C/cm^2$ [10]. The magnetic order is given by a frustrated triangular basal plane spin structure [11]. Nénert et al.[12] revealed a new intermediate phase between 1125 K and 1350 K in single crystal of $YMnO_3$ using a dilatometry and differential thermal analysis. The existence of the intermediate phase is still under debate, but on the other hand there is relatively large spread of published $T_C$ values in literature from 913 K to 1270 K [9]. Fennie and Rabe [13] made first principle calculations and came to the conclusion that the intermediate phase is energetically less favourable than a single improper ferroelectric (FE) phase transition. According to them the FE phase transition is caused by the softening of an optic phonon at the Brillouin zone (BZ) boundary with **q** = (1/3, 1/3, 0) and spontaneous polarization arises from the coupling of this phonon to a zone center mode. If this is true, no pronounced dielectric anomaly is expected near $T_C$. However, direct dielectric measurements are missing near $T_C$ due to a rather high conductivity above room temperature. We found just one old conference paper [5] about dielectric studies of $YMnO_3$ performed at 150 MHz up to 1050 K showing a relatively small permittivity peak and large dielectric loss near 910 K.

In contrast to hexagonal $YMnO_3$, the ferroelectricity in orthorhombic-$RMnO_3$ occurs below the magnetic phase transition (typically below 40 K) due to locking of the incommensurate helicoidal spin structure to a commensurate one leading to the creation of transverse polarization as a consequence of interaction between the spins. Due to such spin-induced ferroelectricity, the magneto-electric coupling is very large in the orthorhombic $RMnO_3$ [14].

In the case of hexagonal $YMnO_3$ the magneto-electric coupling has different origin. Linear magnetoelectric coupling is forbidden by symmetry, but it was shown that the antiferromagnetic domain walls strongly interact with the lattice strain in ferroelectric domain walls due to the piezomagnetic effect [1,15], so any switching of ferroelectric polarization triggers flipping of the antiferromagnetic order parameter. Very recently Choi et al. [16] discovered that the ferroelectric domain walls and structural antiphase boundaries are mutually locked and this strong locking leads to an incomplete poling even under a high electric field.

In hexagonal $YMnO_3$ the spin-phonon coupling is anomalously large. This was demonstrated in X-ray diffraction studies [17], dielectric permittivity [18,19,20], thermal conductivity [21] and in inelastic neutron scattering studies [22]. Infrared (IR) [23] and Raman scattering spectra [24] revealed a pronounced phonon hardening near and below $T_N$ due to spin-lattice coupling, which is responsible for the decrease of dielectric permittivity [19] near $T_N$.

Inelastic neutron scattering (INS) revealed three magnon branches. Two of them are degenerate near the BZ centre and have frequencies near 40 $cm^{-1}$ [25,26]. The AFM resonance was confirmed at the same frequency by means of FIR spectroscopy [27]. Recent polarized inelastic neutron scattering studies revealed that the excitation seen at 1.5 K near 40 $cm^{-1}$ has a hybrid character of magnetic spin wave and a lattice vibration [28]. In other words, it should contribute to both magnetic permeability and dielectric permittivity. The authors proposed to explain this mode hybridization by Dzyaloshinskii – Moriya interaction [28].

In this paper we report on THz and IR studies of hexagonal $Y_{1-x}Eu_xMnO_3$, x=0, 0.1 and 0.2 ceramics. Temperature and Eu concentration dependence of AFM resonance observed near 40 $cm^{-1}$ was determined. The IR reflectivity studies of $YMnO_3$ single crystal below 300 K have been recently published [23]. In this paper we extend these investigations up to 900 K, i.e. close to the ferroelectric phase transition; in addition our IR data are complemented by THz measurements for more accurate determination of the permittivity up to 900 K. Based on our data we confirm the improper ferroelectric character of the phase transition, as recently proposed by Fennie and Rabe [13].

**Experimental details**

High quality $Y_{1-x}Eu_xMnO_3$ ceramics were prepared by the sol-gel urea combustion method. A detailed study of $EuMnO_3$ and $GdMnO_3$ ceramics prepared in this way has lead to results very similar to those obtained for the corresponding single crystals [29]. The valence of the europium ion was checked through XPS technique, and no evidences for other valences than the Eu (III) could be detected. As the samples were fast cooled from 1350 ºC down to room temperature, significant deviations of the oxygen occupancy are not expected, excluding the existence of significant amounts of Mn (IV) ions [30].

IR transmission and reflectivity measurements were performed using a Fourier Transform IR Spectrometer Bruker IFS 113v in the temperature range of 5 K ÷ 900 K and frequency range between 15 and 3000 $cm^{-1}$ (0.45 - 90 THz) with the resolution of 2 $cm^{-1}$. A helium-cooled Si bolometer operating at 1.6 K was used as a detector for low-temperature measurements while a pyroelectric DTGS detector was used above room temperature. Time-domain THz spectroscopy measurements were performed in the range of 100 GHz and 2 THz. Linearly polarized THz probing pulses were generated using a Ti:sapphire femtosecond laser whose pulses iluminated an interdigitated photoconducting GaAs switch. THz signal was detected using the electro-optic sampling with 1 mm thick [110] ZnTe crystal. An Optistat CF Oxford Instruments continuous flow helium cryostat was used for cooling the samples down to 10 K and a commercial high temperature cell Specac P/N 5850 was used for heating the samples up to 900 K in both IR and THz spectrometers.

The plane-parallel samples were polished on both sides to obtain a mirror like surface. The thickness and diameter of all $Y_{1-x}Eu_xMnO_3$ ceramic samples were 1.032 mm and 9 mm, respectively.

**Results and discussion**

Figure 1 shows the IR reflectivity spectra of YMnO$_3$ between 10 and 900 K. Spectra are plotted only below 700 cm$^{-1}$, because the spectra at higher frequencies are temperature independent. Spectra of Y$_{1-x}$Eu$_x$MnO$_3$ ceramics (not shown here) are very similar to those of YMnO$_3$. A gradual decrease of reflection band intensities with increasing temperature is seen. This is caused by an increase of the phonon damping with temperature as will be discussed below. Parameters of the observed phonons were obtained from the fits. The reflectivity spectra $R(\omega)$ are related to spectra of the complex index of refraction $n^*(\omega)$ using the relation

$$R(\omega) = \left| \frac{n^*(\omega)-1}{n^*(\omega)+1} \right|^2 \quad \text{where} \quad n^*(\omega) = \sqrt{\varepsilon^*(\omega)\mu^*(\omega)} \, . \quad (1)$$

The complex magnetic permeability $\mu^*(\omega) = \mu'(\omega) + i\mu''(\omega)$ exhibits usually no dispersion in the THz spectra of dielectrics, so $\mu'(\omega) = 1$ and $\mu''(\omega) = 0$ and thus $n^*(\omega) = \sqrt{\varepsilon^*(\omega)}$ in non-magnetic materials. However, in our AFM material some magnetic excitation can appear in the THz or IR spectra and since we cannot unambiguously distinguish between magnetic and dielectric excitations from our spectra of ceramics, we will systematically plot the real and imaginary parts of the product $\varepsilon^*(\omega)\mu^*(\omega)$ spectra.

The complex permittivity $\varepsilon^*(\omega)$ is in our model described by a sum of classical damped quasiharmonic oscillators

$$\varepsilon^*(\omega) = \varepsilon_\infty + \sum_{j=1}^{n} \frac{\Delta\varepsilon_{TOj}\omega_{TOj}^2}{\omega_{TOj}^2 - \omega^2 + i\omega\gamma_{TOj}} \quad (2)$$

where $\omega_{TOj}$, $\Delta\varepsilon_{TOj}$ and $\gamma_{TOj}$ is the eigenfrequency, dielectric strength and damping of the j$^{th}$ polar phonon, respectively. $\varepsilon_\infty$ is the high frequency permittivity due to electronic absorption processes in the visible and UV ranges. The static permittivity $\varepsilon(0)$ is given by

$$\varepsilon(0) = \sum_{j=1}^{n} \Delta\varepsilon_{TOj} + \varepsilon_\infty \quad (3)$$

In classical displacive ferroelectrics, where no dispersion exists below the phonon frequencies, $\varepsilon(0)$ usually corresponds to the low frequency permittivity measured in radiofrequency range. We note that magnetic permeability spectra $\mu^*(\omega)$ can be fitted as well with a harmonic oscillator model similar to Eq.(2).

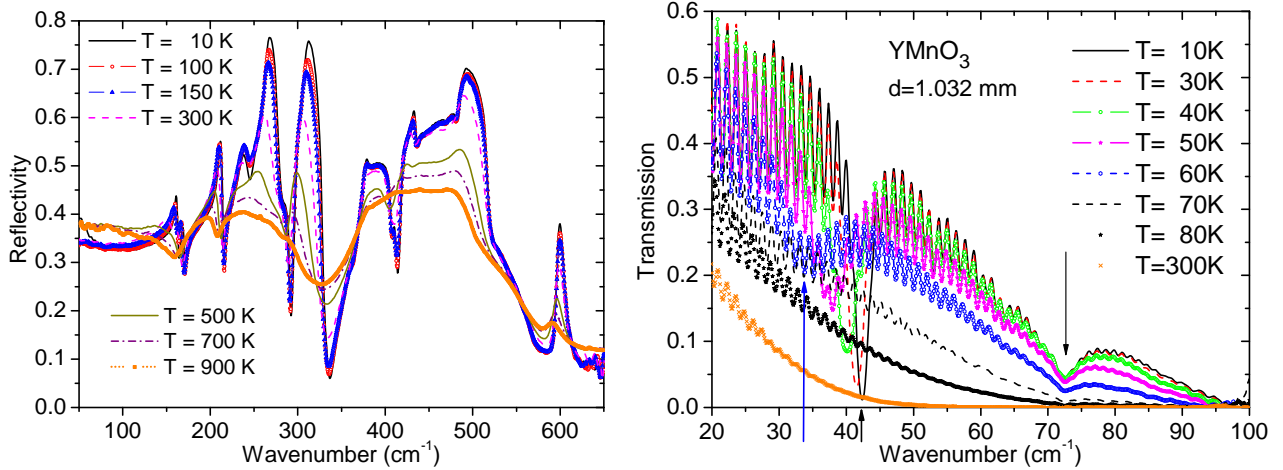

Figure 1. Temperature dependence of the infrared reflectivity spectra of $YMnO_3$ ceramics.

Figure 2. Far-IR transmission spectra at various temperatures. At 10 K, two absorption bands are clearly observed around 42 cm$^{-1}$ and 73 cm$^{-1}$, the former band shifts down to 34 cm$^{-1}$ at 60 K. The frequencies of absorption bands are marked by arrows.

The low-frequency part (below 100 cm$^{-1}$) of our far-infrared (FIR) spectra was actually investigated by means of three independent techniques: 1) FIR reflectivity (Figure 1), FIR transmission (Figure 2) and time-domain THz transmission spectroscopy, which gives directly the real and imaginary parts of the product $\varepsilon^*(\omega)\mu^*(\omega)$ (Figure 3). All these spectra were fitted simultaneously and the obtained complex spectra are shown in Figure 4.

A sharp resonant peak is seen near 40 cm$^{-1}$ in Figures 2 and 3. This is the AFM resonance which was briefly described in ref. [27]. The AFM resonance causes dispersion only in $\mu^*(\omega)$ spectra and its frequency decreases on heating towards $T_N$ (see Figure 5). Similar AFM resonances with slightly lower frequencies were observed also in $Y_{1-x}Eu_xMnO_3$ ceramics (x= 0.1 and 0.2).

Phonon frequencies obtained from the fits of IR reflectivity exhibit no dramatic change with temperature, only a small gradual decrease of their values was observed (see Figure 6). This phonon softening will be explained below.

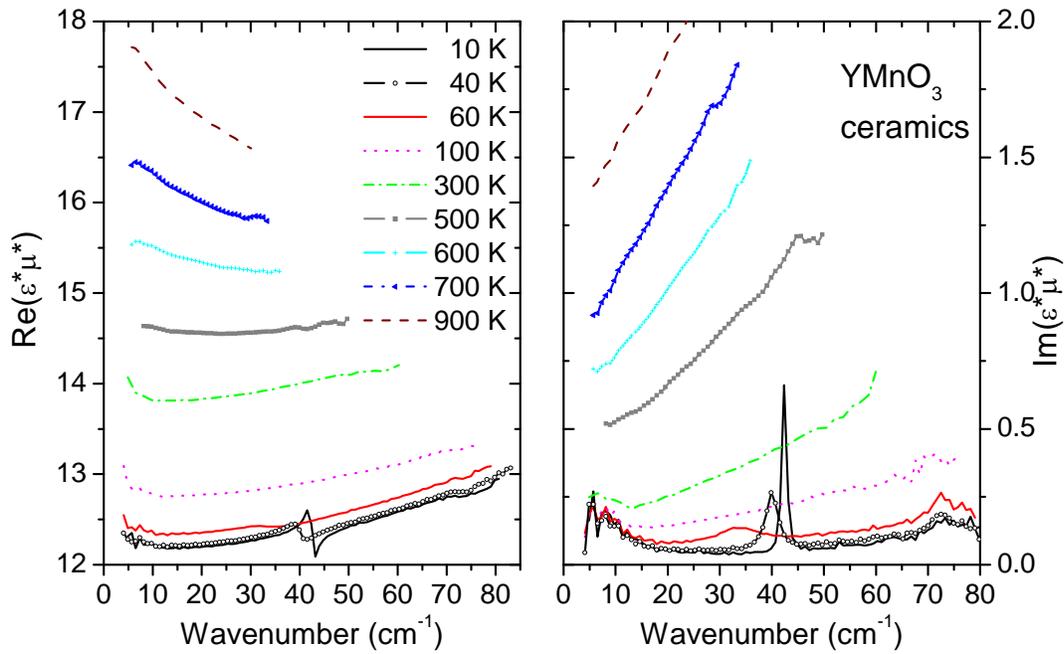

Figure 3. Temperature dependence of real and imaginary parts of experimental THz spectra of the product $\varepsilon^*\mu^*$ in a broad temperature range. The spin wave near 40 cm$^{-1}$ is clearly seen, a weaker excitation near 73 cm$^{-1}$ is also apparent – see also FIR transmission spectra in Figure 2.

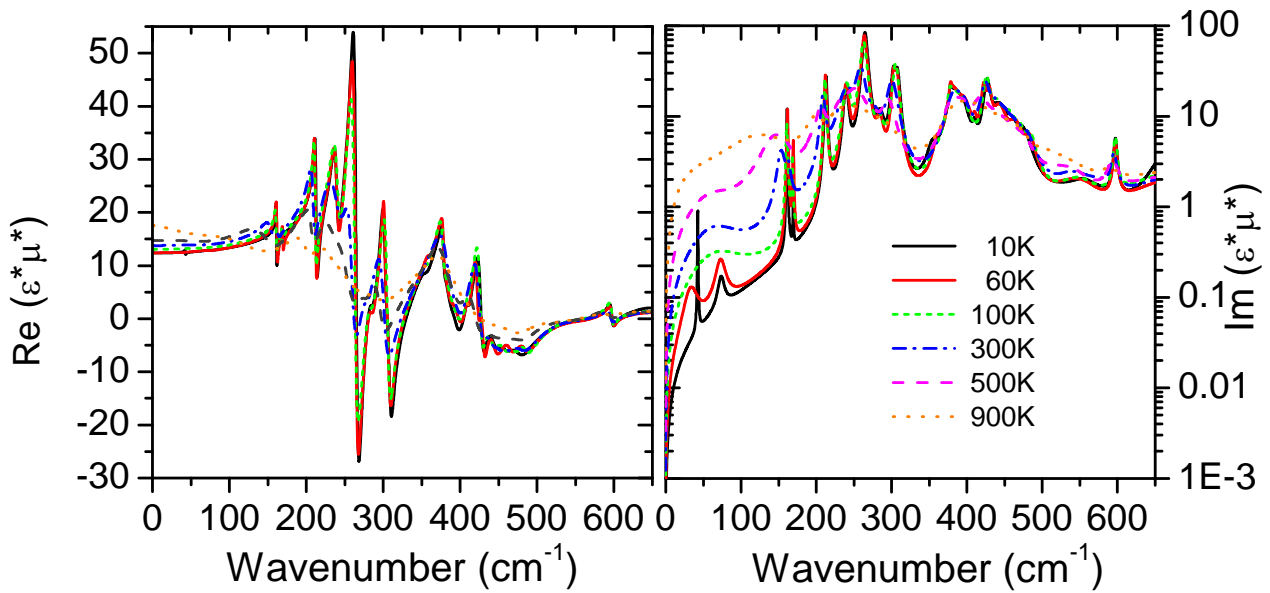

Figure 4. Real and imaginary parts of $\varepsilon^*\mu^*$ in YMnO$_3$ ceramics at various temperatures obtained from the simultaneous fit of THz and FIR spectra. Peaks in the imaginary part correspond to the phonon and magnon frequencies.

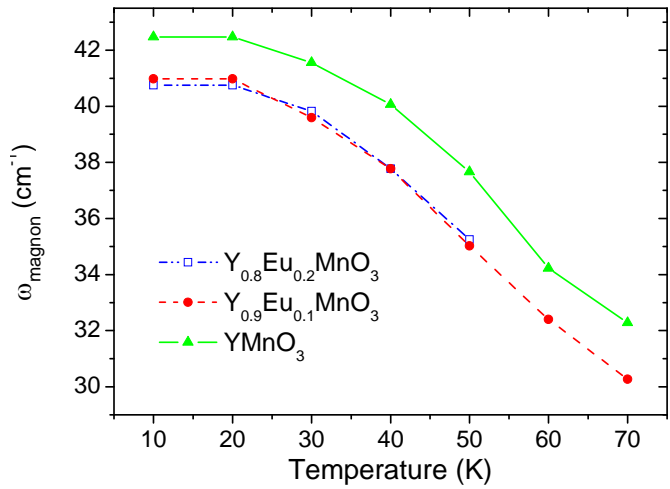 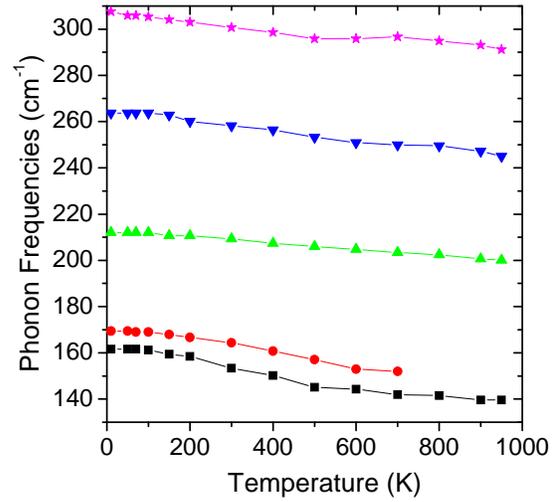

Figure 5. Temperature dependence of the antiferromagnetic resonance frequencies in the $Y_{1-x}Eu_xMnO_3$ ceramics for x=0, 0.1 and 0.2

Figure 6. Temperature dependence of the selected phonon frequencies in the $YMnO_3$ ceramics.

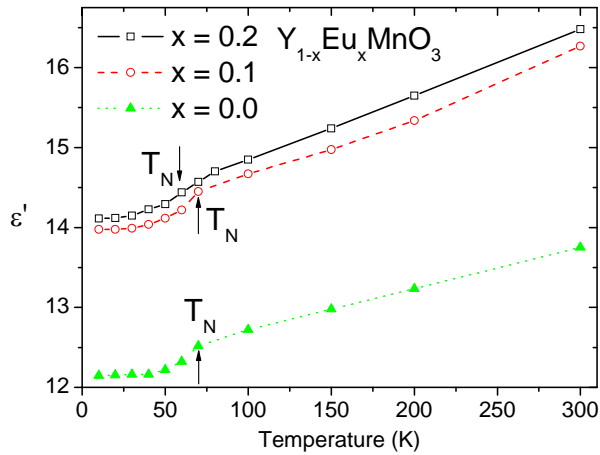

Figure 7. Temperature dependence of ε' in the $Y_{1-x}Eu_xMnO_3$ (x=0, 0.1 and 0.2) ceramics at 0.5 THz from THz spectra. Lines are guides for the eyes.

Polarized IR reflectivity spectra of a $YMnO_3$ single crystal were published by Zaghrioui et al. [23]. 9 $A_1$ and 14 $E_1$ symmetry polar modes are allowed from the factor group analysis [31] for **E** of the IR beam parallel and perpendicular to the *c* crystal axis, respectively. Zaghrioui et al. [23]

observed 7 $A_1$ modes and 8 $E_1$ modes. Our spectra of ceramics shown in Figure 1 are unpolarized, therefore we see a mixture of both polarized spectra. Especially above 350 cm$^{-1}$ the modes are highly overlapped and it is very difficult to distinguish all the modes in the reflectivity spectra. In spite of this we used 24 oscillators for the fits of the spectra at 10 K (including one magnon) which exactly corresponds to number of polar phonons expected from the factor group analysis [31]. Ten polar-mode frequencies have similar values to those observed in the single crystal [23]. The other modes show slightly different frequencies; this is because the frequencies of effective modes in ceramics can be shifted with respect to the pure frequencies of $A_1$ and $E_1$ modes. Figure 6 shows frequencies of phonons which exhibit some more pronounced temperature dependences. Other modes, which are not plotted in Figure 6, show the following frequencies (in cm$^{-1}$) at 10 K: 73.8, 239.5, 284.8, 302.2, 354.4, 378.2, 382.8, 394.7, 410.3, 426.7, 442.3, 457.9, 472.5, 477.1, 555.0, 597.1, and 666.7. The small decrease of phonon frequencies observed on heating in Figure 6 is mainly due to the thermal expansion with increasing temperature. The phonon damping increases with rising temperature as well, therefore the intensity of reflectivity bands in Figure 1 decreases at high temperatures.

The THz spectra taken above $T_N$ show the evolution of the complex permittivity spectra, as the AFM resonance contributing to $\mu^*(\omega)$ only exits below $T_N$. One can see that the real part of the permittivity ($\varepsilon'$) increases on heating so that the static permittivity $\varepsilon'(0)$ goes from 12.5 at 10 K up to 17.8 at 900 K. Such a change is rather small and is caused by the small phonon softening seen in Figure 6. In the case of a proper ferroelectric phase transition $\varepsilon'(0)$ should raise on heating towards $T_C$ according to the Curie-Weiss law. Our observation of a less pronounced permittivity increase on heating supports the improper character of the ferroelectric phase transition originally suggested by Fennie and Rabe [13]. In this case the ferroelectric phase transition is driven by a soft phonon from the Brillouin zone boundary with **q**=(1/3, 1/3, 0), which does not contribute to the permittivity above $T_C$. It can contribute to $\varepsilon'(0)$ only in the ferroelectric phase, but the dielectric strength of this newly IR activated mode is usually very small, so the change of $\varepsilon'(0)$ with temperature is also very weak. A slight change of the THz permittivity near $T_C$ can also be expected in the case of an order-disorder phase transition, but then a larger structural disorder should be observed in paraelectric phase, which did not occur [8,9].

In the literature there are several reports about an intermediate phase between ferroelectric and paraelectric phases [8,12,32]. Fennie and Rabe [13] have shown, based on symmetry analysis, that the intermediate phase, if it exists, should be paraelectric $P6_3/mcm$ (Z=6). It means that, within this hypothesis, the ferroelectric phase transition should be a proper ferroelectric one driven by a zone-center polar soft mode and a large dielectric anomaly at $T_C$ would be expected. Fennie and Rabe have shown that this scenario is energetically less probable than the improper ferroelectric phase transition and our small experimental permittivity seen 40 K below $T_C$ really supports the improper character of the transition. Nevertheless, one must admit that the weak permittivity increase could also occur if $T_C$ were much higher than 900 K. This is less probable, but some experimental papers [8,12] reported about it.

FIR transmission (Figure 2) and THz spectra (Figure 3) reveal a sharp resonant absorption near 40 cm$^{-1}$ in the AFM phase of YMnO$_3$. This is the AFM resonance which was briefly reported in Ref. 27 and its frequency corresponds also to a zone-center magnon seen in INS spectra [22,25,26,28]. The same AFM resonance was revealed also in the THz spectra of $Y_{1-x}Eu_xMnO_3$ with x=0.1 and 0.2; its frequency is just slightly lower than for YMnO$_3$ (see Figure 5).

Another sharp absorption is seen in the FIR transmission spectra near 73 cm$^{-1}$ (Figure 2). Its signature is observed in the THz spectra too (Figure 3). Unfortunately, it can only be observed at low temperatures below 60 K, i.e. in the AFM phase, because the sample is not sufficiently transparent at higher temperatures. From these data it is not clear whether this excitation has a magnetic or polar origin. This mode is not directly observed in IR reflectivity spectra, because its strength is very small. However, a simultaneous fit of the IR reflectivity and THz spectra above $T_N$ requires this heavily damped mode up to 900 K. Its existence in the paramagnetic phase suggests its polar origin. We stress that without this mode it is impossible to fit the enhanced THz dielectric losses in Figure 3. It can be seen in Figure 4 as a broad loss maximum below 100 cm$^{-1}$. Its oscillator strength increases on heating, which is not typical for a single phonon absorption process, but is possible for a multiphonon one. Nevertheless, multiphonon excitation should die out on cooling. In our case we see this mode down to 10 K, which suggests its possible defect origin. This assumption is also supported with our preliminary IR transmission and THz spectra of YMnO$_3$ single crystal, where this mode was not observed.

Temperature dependence of the permittivity obtained in $Y_{1-x}Eu_xMnO_3$ at 0.5 THz shows a small drop near $T_N$ (see Figure 7) which is a consequence of a large spin-lattice coupling responsible for an isostructural phase transition at $T_N$ and for the atomic displacements revealed in the XRD studies [17]. The spin-lattice coupling causes hardening of some phonons near $T_N$, especially of the $E_1$-symmetry mode observed in the IR spectra of a single crystal [23]. Similar dielectric anomaly was observed in the 1 MHz data published in [19] with a slightly higher value of ε'. This was probably caused by a higher density of the ceramic sample investigated in [19]. The temperature where the drop in ε' appears can be used for a rough estimation of $T_N$. One can see that $T_N$ is approximately the same in $Y_{1-x}Eu_xMnO_3$ for x=0, 0.1 and 0.2 (around 70K as usually reported in the literature). Nevertheless, in x=0.2 sample the dielectric anomaly is more smeared than in the rest of ceramics, giving evidence of broadening of the AFM phase transition with Eu substitution.

# Conclusion

High-temperature IR reflectivity spectra of $YMnO_3$ reveal a small gradual decrease of several phonon frequencies, which is responsible for the observed weak increase of the THz permittivity on heating towards 900 K. Such phonon and dielectric behavior supports the improper character of the ferroelectric phase transition near 940 K connected with a tripling of the unit cell and with the soft mode anomaly at the Brillouin zone boundary. Our experimental data support the theoretical predictions of Fennie and Rabe [13] and do not confirm the existence of an intermediate paraelectric phase with tripled unit cell above $T_C$ suggested in papers [8,12]. The apparent existence of the intermediate phase is probably caused by defects and related to the diffuse character of the phase transition.

A sharp AFM resonance and the softening of its frequency on heating towards $T_N$ was revealed in the THz spectra. A strong spin-phonon coupling is demonstrated by a significant decrease of the THz permittivity below $T_N$. Eu substituted $YMnO_3$ exhibits magnetic and dielectric properties similar to those of $YMnO_3$.

**Acknowledgment**

This work was supported by Fundação para a Ciência e Tecnologia, through the Project PTDC/CTM/67575/2006, and by Programme Alban, the European Union Programme of High Level Scholarships for Latin America (Scholarship no. E06D100894BR). This work was also supported by the Czech Science Foundation (Project 202/09/0682), SVV-2010-261303 and AVOZ 10100520. The authors thank J. Petzelt for critical reading of the manuscript.